# Impurity state in Haldane gap for $S = 1$ Heisenberg antiferromagnetic chain with bond doping

Xiaoqun Wang and Steffen Mallwitz
*Max-Planck-Institut für Physik Komplexer Systeme, Bayreuther Str. 40, Haus 16, 01187 Dresden, Germany*

Using a new impurity density matrix renormalization group scheme, we establish a reliable picture of how the low lying energy levels of a $S = 1$ Heisenberg antiferromagnetic chain change *quantitatively* upon bond doping. A new impurity state gradually occurs in the Haldane gap as $J' < J$, while it appears only if $J'/J > \gamma_c$ with $1/\gamma_c = 0.708$ as $J' > J$. The system is non-perturbative as $1 \leq J'/J \leq \gamma_c$. This explains the appearance of a new state in the Haldane gap in a recent experiment on $Y_{2-x}Ca_xBaNiO_5$ [J.F. DiTusa, *et al.*, Phys. Rev. Lett. **73** 1857(1994)].

Since Haldane [1] predicted that integer-spin antiferromagnetic (AF) chains have a gap in the excitation spectrum, much work [2,3] has been done on the $S = 1$ Heisenberg AF chain. In particular, the valence bond solid (VBS) model proposed by Affleck *et al* [3] has succeeded in verifying this prediction. Recently, very accurate estimates of the Haldane gap ($\Delta = 0.41049(2)J$) have been achieved by White and Huse [4] and Golinelli, *et al* [5], using the density matrix renormalization group (DMRG) method and exact diagonalization together with proper extrapolations, respectively. Evidence for the gap has also been clearly shown by some experiments on quasi-1D $S = 1$ AF materials: NENP [6] and $Y_2BaNiO_5$ [7–9]. Impurity effects have not been fully studied yet, although the VBS picture indicates that the physics of low lying excitations could be dramatically affected by the introduction of impurities. ESR on NENP [10] shows evidence for the existence of free $S = 1/2$ spins at the chain ends upon magnetic doping with Cu or non-magnetic doping with Zn, Cd and Hg, whereas the specific heat measurements on $Y_2BaNi_{1-x}Zn_xO_5$ show a Schottky anomaly consistent with free $S = 1$ spins [9]. Recently DiTusa *et al* [8] have performed an experiment with carrier doped $Y_{2-x}Ca_xBaNiO_5$. The inelastic neutron scattering shows that the Ca doping produces *new states* in the Haldane gap. According to polarized x-ray absorption spectra, the added hole is localized on the oxygen site with $S = 1/2$ spin, which might either directly break the exchange coupling $J$ and thus couple to $S = 1$ $Ni^{+2}$ ions or otherwise change $J$ into $J'$. It is clear that the impurity effects are fundamental for further understanding the properties of Haldane gap systems.

Recently, Sørensen and Affleck [10] have studied the low-lying excitations in these doped Haldane systems using the DMRG method [11] and a free massive field theory. One of the most surprising findings is that critical values for $J'$ are required to induce impurity states in the Haldane gap both in the weak and strong coupling regimes. However, as remarked by the authors, their calculation using the standard DMRG method becomes progressively inaccurate for small or large couplings, since the system is split *exactly* at the impurity bond and correlation effects of the impurity bond entering truncated blocks are enhanced as the chain grows. Their results do also not agree with those obtained by a Schwinger boson technique [13], which gives rise to an impurity state in the middle of the Haldane gap being almost independent of $J'$. Therefore, a substantially more accurate calculation is needed to clarify the fundamental properties of such doped Haldane systems.

The DMRG technique [12] invented by White has successfully provided highly accurate results for some homogeneous systems. For a given number of states kept in the truncation procedure, numerical results have a higher precision using open rather than periodic boundary conditions. On the other hand, restricting oneself to open boundary conditions leads to an unfavourable scheme for the application of DMRG to impurity systems. One can nevertheless provide very accurate results even for impurity systems if one is able to work with *a periodic boundary* and *sufficiently many states. Periodic boundary conditions* allow for a proper placement of the impurities in the chain. A general scheme suitable for an impurity problem (called the impurity scheme hereafter) is shown in Fig. 1. In this scheme, a truncation is carried out only for the blocks B so that the impurity part contained in the dashed box is not subject to the truncation procedure [15]. Moreover, with increasing chain length, added sites depart further and further from the impurity part, which is in the spirit of Wilson's renormalization group as applied to the Kondo impurity problem in momentum space. In order to keep *sufficiently many states*, a well-optimized code is designed, which is suitable for quite general DMRG calculations.

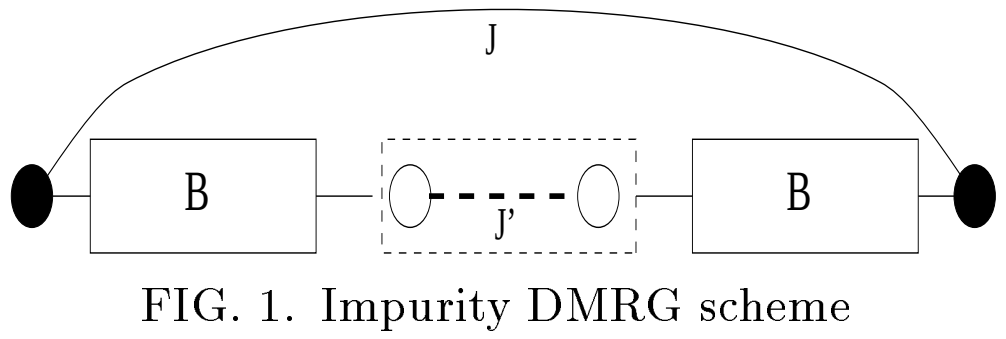

FIG. 1. Impurity DMRG scheme

In this letter, we consider the case where due to doping one bond $J$ of a closed $S = 1$ Heisenberg AF chain is changed to $J'$. The corresponding Hamiltonian is

$$H = J \sum_{i=1}^{L-1} \mathbf{S}_i \cdot \mathbf{S}_{i+1} + J' \mathbf{S}_1 \cdot \mathbf{S}_L \qquad (1)$$

where $J > 0$ and $0 \leq J' \leq \infty$. It is essential to point out some relevant properties of eq. (1) for special cases in order to understand our numerical results: i)For $\gamma = J'/J = 1$, eq.(1) reduces to a $S = 1$ Heisenberg AF chain of length $L$ [2]. For even $L$, the ground state is a singlet. In the thermodynamic limit continuous excitations start at the Haldane gap and the lowest one is a triplet state with $S = 1$ and $\mathbf{q} = \pi$. ii)As $\gamma = 0$, i.e. the open boundary case [14], the lowest excitation with $S = 1$ becomes degenerate with the singlet ground state in the thermodynamic limit. The Haldane gap still exists, but is now the energy difference between the ground states and the $S = 2$ states. In this case, the chain end spins are fully polarized. iii)As $\gamma \to \infty$, the spins linked by the impurity bond form a singlet, which leaves the problem of a $L-2$ sites chain with an open boundary. All properties of the open chain are present apart from this isolated singlet. As long as $\gamma \neq 0, 1, \infty$, we believe that the lowest excitations are rearranged and an impurity state is associated with the lowest excited state with $S = 1$. In the following, we employ exact diagonalization and DMRG methods together with proper extrapolations to clarify how the triplet state descends from the bottom of the continuum spectrum at $\gamma = 1$ to the ground state at $\gamma = 0, \infty$ and whether the properties of the system in the case of $\gamma > 1$ are the same as for $\gamma < 1$.

Now we turn to some important details of our numerical calculation for impurity energy and more details will be given elsewhere. We use the standard DMRG as $\gamma = 0, 1$, for which 500 and 1100 states are kept in each block B, respectively. In the impurity scheme, we conduct the calculation exactly up to the 16 site chain which contains 729 states in each block B, and subsequently we keep 600-630 states for the remaining calculation. For $\gamma = 1$, we found that the impurity scheme keeping 600 states gives a result of similar accuracy as that given by the standard DMRG scheme using periodic boundary and keeping 1100 states. As $\gamma \neq 0, 1$, we also checked how reliable the standard DMRG results are using either open or periodic boundary. We found that the inaccuracy becomes apparent as $\gamma$ is either small or large, whereas the impurity scheme gives very accurate results even for very small and very large values of $\gamma$. Up to 40 site-chains [16], the truncation error is at most $7.0 \times 10^{-9}$ and typically of the order of $10^{-10}$.

To determine the accuracy of our calculation, we checked it by keeping different numbers of states. The systematic error of the energies due to truncation is smaller or of the order of $10^{-6}$. Moreover, exact results for $\gamma = 1$ are available for up to 22 sites, according to which the difference to DMRG results with 1100 states kept are $4.0 \times 10^{-9}$ and $4.3 \times 10^{-8}$ for ground state energy and Haldane gap, respectively. We obtained $\Delta = 0.410499$ from DMRG of 22 sites with Shanks tranformation, which is essentially the same as that given by exact diagonalization with Shanks transform. [5] To obtain final results we eliminate finite size effects for each $\gamma$ by both fitting all data to exponential laws and using Shanks transform with smaller chains ($L = 22 \sim 26$). We found that these two results coincide very well. The errors of the final results we obtained for the impurity state are at most in the fourth digit as is partially shown in Table I, which we consider sufficiently accurate to ensure the reliability of our conclusions.

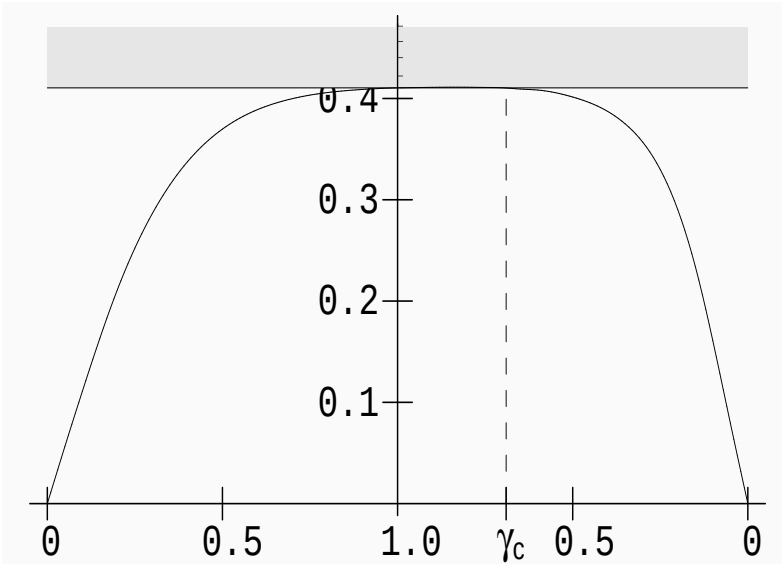

FIG. 2. Energy versus $J'/J$(left) and $J/J'$(right)

The dependence of the energy of the low-lying states on bond doping is given in Fig. 2. The horizontal axis indicates the ground-state energy. The darker shaded area represents the continuum spectrum and the straight line at its bottom marks the Haldane gap which is independent of the impurity coupling $J'$. According to the analysis of finite chains of up to 16 sites, we found that the bottom state has $S = 2$ at $\gamma = 0, \infty$ and $S = 1$ otherwise. Using DMRG, we checked $\Delta = 0.411$ for $\gamma = 0.6$ [17]. Solid curves represent the impurity state, whose appearance in the Haldane gap depends on the strength of the impurity coupling $\gamma J$. In the weak coupling regime shown on the left side, it appears as long as $0 < \gamma < 1$, whereas in the strong coupling regime shown on the right side it does only if $\gamma_c < \gamma < \infty$.

To interpret the above results, we show some detailed data in Figs. 3 and 4 and insets show amplified critical regimes. For any finite $L$, the impurity state has $S = 1$ forming a degenerate triplet. In order to characterize this state, we calculated the energy difference $\Delta^L(\gamma)$ between the ground state and the triplet state. The energy of this triplet state can be evaluated as the lowest energy in the $S_z = 1$ subspace. In Fig. 3 $\Delta^L(\gamma)$ is denoted by an open square for given $L$ and $\gamma$. The values of $\Delta^L(\gamma)$ decrease from top to bottom as $L$ increases for given $\gamma$. For each

$L \in [8, 22]$ a fitting curve links these squares as a guide

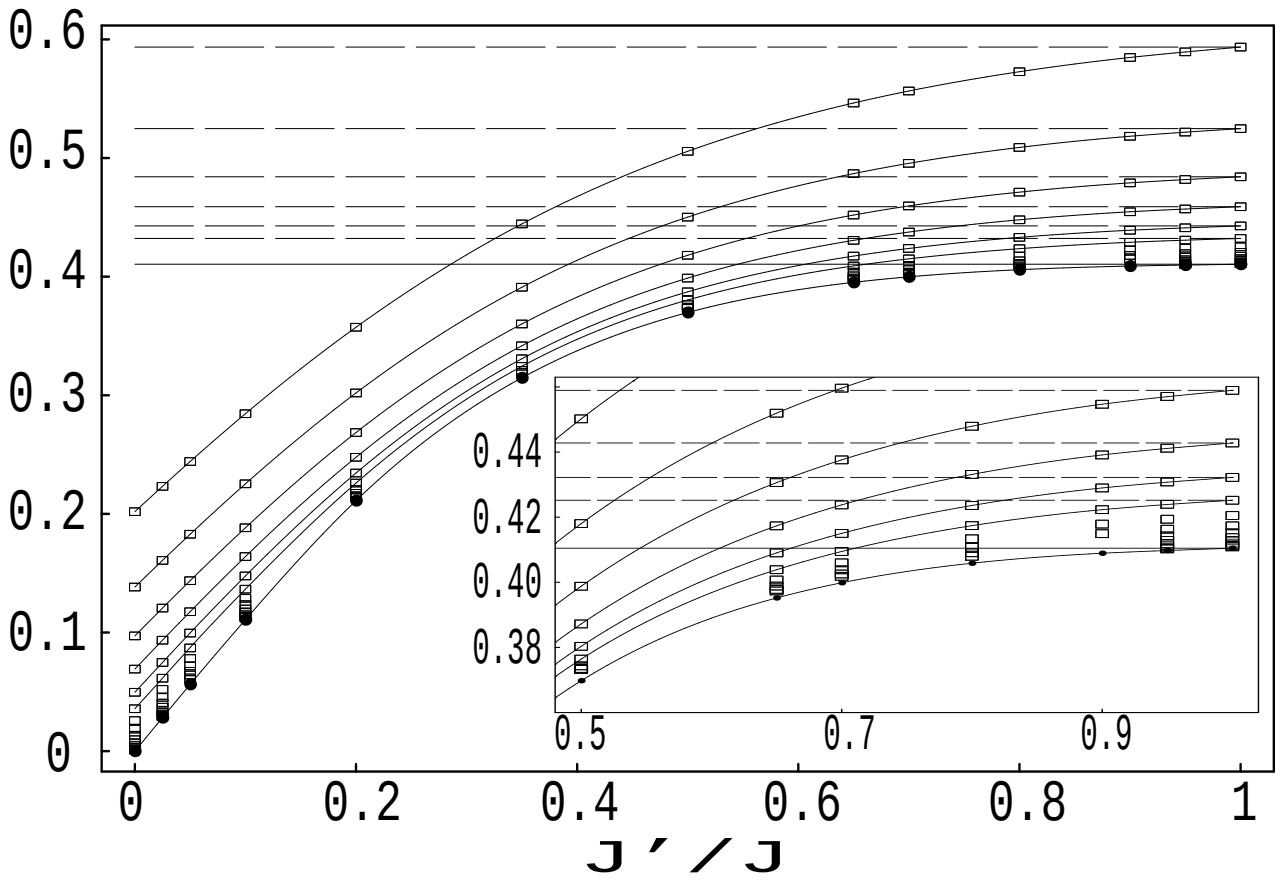

FIG. 3. Weak coupling region

to the eye and dashed lines denote $\Delta^L(1)$. As is explicitly shown in Fig. 3, we find

$$\Delta^L(\gamma < 1) < \Delta^L(1).$$

This behaviour certainly persists as $L \to \infty$. Using the extrapolation discussed above, we obtain $\Delta^\infty(\gamma)$ as denoted by solid dots, which is fitted to a solid curve. This curve is below the Haldane gap denoted by a solid straight line. However, the situation is completely different for $\gamma \geq 1$ as shown in Fig. 4. For any finite $L$, there exists an overshooting and the fitting curve intersects twice the dashed line which indicates the Haldane gap at that particular $L$. It is clearly seen that

$$\Delta^L(\gamma_c^L < \gamma) < \Delta^L(\gamma = 1) \leq \Delta^L(1 \leq \gamma \leq \gamma_c^L),$$

where $\gamma_c^L$ is the value of $\gamma$ corresponding to the intersection point for given $L$. To estimate $\gamma_c$ for the critical point, we first utilize an extrapolation of the intersection points $\gamma_c^L$. Secondly we determine the point at which the fitted curve of $\Delta^\infty(\gamma)$ drops below the Haldane gap. These two independent estimates yield $1/\gamma_c = 0.708(2)$ where the error shows their difference.

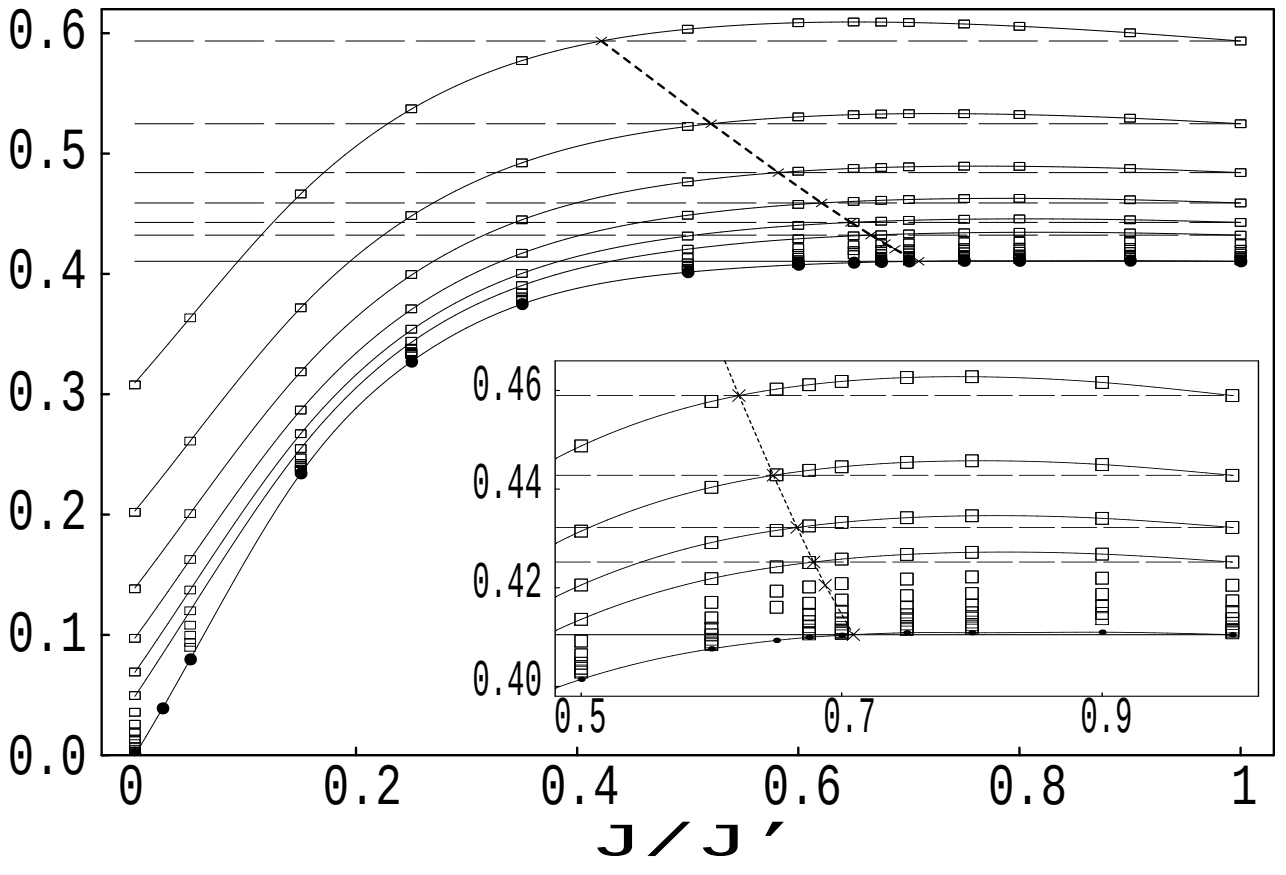

FIG. 4. Strong coupling region

In order to further investigate the properties of such a doped system, one can map the original problem onto a $L-2$ system by projecting out the singlet spins as $\gamma \gg 1$ [11]. The leading term in $1/\gamma$ is still an impurity bond of a strength $4J/3\gamma$ and other terms are $O(1/\gamma^2)$. Such a mapping certainly breaks down as $\gamma \leq 4/3$, where either higher orders or the break up of the singlet become crucial, resulting in significantly different properties for the strong coupling as compared to the weak coupling regime. Therefore, according to our numerical results, the properties of such a doped system are non-perturbative as $1 \leq \gamma \leq \gamma_c$, otherwise the problem can be treated properly in $\gamma$ or $1/\gamma$ in the weak or strong coupling regime, respectively.

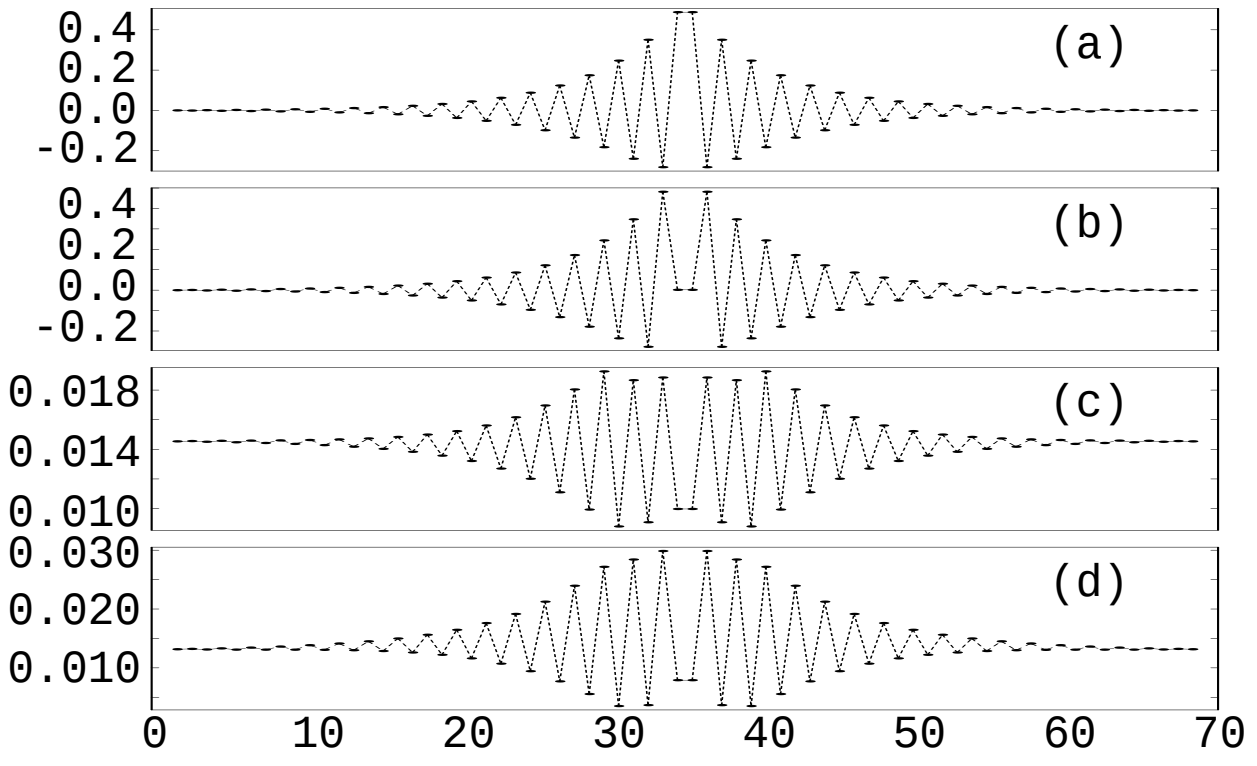

FIG. 5. Spin polarization $P_i$

Furthermore, it is instructive to analyze the spin polarization induced by $J' \neq J$. In this case, we calculate $P_i =< 1|S^z_{i+35}|1 >$, i.e. the expectation values of $S^z_{i+35}$ in the state with $S_z = 1$, and the impurity bond is translated to the center of the periodic chain. As $\gamma = 1$, we have $P_i = 1/L$. In Fig. 5, we show $P_i$ for four values of $\gamma$ with $L = 70$ and 300 states kept. In the weak and strong coupling limits, the $P_i$ are essentially the same apart from a singlet in the middle as shown in (a) and (b) for $\gamma = 0.1, 40/3$, respectively. As $\gamma = 1/0.8$ which is in the non-perturbative regime, the region of sizeable polarization (c) is still similar to the cases (a) and (b). On the other hand, we find for (a) and (b) $\chi(\gamma, i) = lim_{L\to\infty} L\,P(i)$ very small for $i \gg 1$ while in the critical regime (c) this limit gives 1. In this sense we can distinguish localised from delocalised states. As can be seen from (d) for $\gamma = 1/0.6$, $\chi(\gamma, i \gg 1)$ is again smaller than 1 but the change is smooth. We believe that $\chi(\gamma, i \gg 1)$ behaves analogous to $\Delta^\infty(\gamma)$, which may give an independent way to determine the critical point.

Now we discuss the properties of such a doped system as $\gamma \to \gamma_w$ and $\gamma_s$ with $\gamma_w = 0, \infty$ and $\gamma_s = 1^-, \gamma_c^+$, respectively. Based on the above analysis and perturbation theory for $\gamma \to \gamma_w$, one finds

$$\Delta^\infty(\gamma) = \begin{cases} \alpha_w J\gamma + O(\gamma^2) & \gamma \ll 1 \\ \alpha_s J/\gamma + O(1/\gamma^2) & \gamma \gg 1. \end{cases}$$

We extracted $\alpha_w = 1.1337$ from $\gamma = 0.025, 0.05$ and $\alpha_s = 1.5229$ from $\gamma = 20, 40$, which gives $\alpha_s/\alpha_w = 1.3433$. The deviation from 4/3 reflects higher order contributions in $\gamma$ or $1/\gamma$. On the other hand, according to the VBS picture, a free $S = 1/2$ edge state exists as $\gamma = 0$. One can project the two spins linked by the impurity bond onto a $S = 1/2$ respresentation [11], giving $\Delta^\infty(\gamma \ll 1) = \alpha^2 J\gamma$ with $< 1|S_1^z|1 >=< 1|S_L^z|1 >= \alpha/2$ as $\gamma \ll 1$. Consequently one has $\alpha^2 = \alpha_w$, which gives rise to $< 1|S_1^z|1 >= 0.5324$ being essentially the same as that obtained by previous calculations under open boundary conditions [12,18]. Regarding the properties of the system near $\gamma_s$, one may assume a power law such as $\Delta^\infty(\gamma) = \Delta + \alpha_n J(\gamma_s - \gamma)^n$ as $\gamma \to \gamma_s$ [11]. According to the perturbation theory with respect to $J' - J$ for $\gamma_s = 1^-$, we can exclude $n = 1$ since $\alpha_1 = \Delta^L(\gamma = 1)/L$. By subtracting this term, we obtain $\alpha_2 = 0.0417$ with $n = 2$ using the results for $\gamma = 0.95$ and $L = 40$. Fitting the data to a polynomial in $1/L$, we have $\alpha_2 \sim 0.000(4)$ being zero within the error of the calculation, so that $n \geq 3$. Clearly *both* more analytical *and* numerical work are needed in order to completely determine this critical behavior.

In conclusion, we have quantitatively established a reliable picture of the appearence of the impurity state in the bond doped Haldane system concerning the off-chain doping material $Y_{2-x}Ca_xBaNiO_5$, where a new state occurs in the Haldane gap. In distinction to Ref. [11,13], we found that in the weak coupling regime, an infinitely small deviation of $J'$ from $J$ gives rise to such a triplet state, whereas in the strong coupling regime, the system exhibits as a fundamental difference not only the formation of a singlet for the spins linked by the impurity bond, but also the existence of a critical point. As $1 \leq \gamma \leq \gamma_c$, the system is clearly non-perturbative.

It is a great pleasure to thank P. Fulde, O. Golinelli, Y.M. Li, H. Scherrer, Z.B. Su, T. Xiang, and L. Yu for helpful discussions.

TABLE I. Singlet-Triplet energy difference as function of $\gamma$

| $\gamma$ | $\Delta^L(\gamma)$ | L | Shanks | fit |
|---|---|---|---|---|
| 0.025 | 0.02909114 | 40 | 0.02822 | 0.02824 |
| 0.050 | 0.05702284 | 40 | 0.05629 | 0.05629 |
| 0.900 | 0.41493434 | 24 | 0.40899 | 0.40894 |
| 0.950 | 0.41022756 | 40 | 0.40989 | 0.40996 |
| 1.000 | 0.41084535 | 40 | 0.41050 | 0.41048 |
| 1/0.700 | 0.41071869 | 40 | 0.41035 | 0.41037 |
| 1/0.675 | 0.41063334 | 36 | 0.40995 | 0.40999 |
| 1/0.050 | 0.08320159 | 32 | 0.07964 | 0.07968 |
| 1/0.025 | 0.04007422 | 40 | 0.03896 | 0.03888 |